\documentclass[10pt]{iopart}

\usepackage{caption}
\usepackage{graphicx,subcaption}
\usepackage[margin=2cm]{geometry}
\usepackage{bm}
\usepackage[table,dvipsnames]{xcolor}
 
\newcommand{\red}{\textcolor{Black}}

\begin{document}

\title{Unsupervised Clustering and Performance Prediction of Vortex Wakes from Bio-inspired Propulsors}

\author{Alejandro G. Calvet, Mukul Dave, Jennifer A. Franck}\address{Department of Engineering Physics, University of Wisconsin-Madison, Madison, WI, USA.}\ead{jafranck@wisc.edu}

\begin{abstract}
An unsupervised machine learning strategy is developed to automatically cluster the vortex wakes of bio-inspired propulsors into groups of similar propulsive thrust and efficiency metrics. A pitching and heaving foil is simulated via computational fluid dynamics with $121$ unique kinematics by varying the frequency, heaving amplitude, and pitching amplitude. A Reynolds averaged Navier-Stokes (RANS) model is employed to simulate the flow over the oscillating foils at $Re=10^6$, computing the propulsive efficiency, thrust coefficient and the unsteady vorticity wake signature. 
Using a pairwise Pearson correlation it is found that the Strouhal number most strongly influences the thrust coefficient, whereas the relative angle of attack, defined by both the mid-stroke and maximum have the most significant impact on propulsive efficiency. 
Next, the various kinematics are automatically clustered into distinct groups exclusively using the vorticity footprint in the wake. A convolutional autoencoder is developed to reduce vortex wake images to their most significant features, and a k-means++ algorithm performs the clustering. The results are assessed by comparing clusters to a thrust versus propulsive efficiency map, which confirms that wakes of similar performance metrics are successfully clustered together. This automated clustering has the potential to identify complex vorticity patterns in the wake and modes of propulsion not easily discerned from traditional classification methods.
\end{abstract}

%
\vspace{2pc}

\section{Introduction} \label{sec:intro}

Swimming and flying animals rely on the fluid around them to provide lift and/or thrust forces, leaving behind a distinct trail of vorticity. The structure and size of the vortex wake is a blueprint of the animal's kinematic trajectory, holding information about the forces and also the size, speed and direction of motion. This is true for a wide range of scales, from low to high Reynolds number swimmers and flyers, and includes man-made bio-inspired propulsive vessels or the vortex-induced vibration motion of bluff-bodied systems. Due to the diverse range of scales, kinematic motions, and varying objectives of organisms (e.g. speed vs. agility), the current classification of vortex wake modes available in literature is not inclusive. Rather than relying on predefined classifications of wake structure, this paper presents an unsupervised clustering process that automatically distinguishes between wake structures, and successfully groups them into regimes of similar propulsive performance. 

The canonical bluff-body wake characterization is based on the model system of flow over an oscillating cylinder, whose wake properties are described and classified by Williamson and Roshko \cite{Williamson1988}. Using various amplitudes and frequencies of oscillation, their findings show regimes described by ``$mS + nP$", where $m$ is the number of single vortices ($S$) per cycles, and $n$ is the number of clockwise/counter-clockwise pairs ($P$). The characterization is expanded to other systems, including that of oscillating foils,
such as that done by Schnipper et al. \cite{Schnipper2009} for the wake behind a pitching foil at Reynolds numbers of 220-440. The wake patterns were mapped based on pitching amplitude and non-dimensional frequency. 
Transition from a drag producing mode to a thrust generation mode was shown to be associated with a clear \textit{reverse} von K\'arm\'an wake, where two alternating vortices are shed in a $2S$ mode with a counter-clockwise vortex at the top of the stoke and a clockwise vortex at the bottom.

The work by Schnipper et al. is preceded by important experimental findings such as   
 a low-amplitude pitching foil experiment by Koochesfahani \cite{Koochesfahani1989} which demonstrates that a combination of frequency and amplitude are responsible for specific vortex patterns and regimes of momentum excess or deficit.
A similar observation is made by Lai and Platzer \cite{Lai1999} for experiments on a plunging (heaving) foil in which the non-dimensional plunging velocity, a function of both amplitude and frequency, determined the threshold for thrust-generation.
Connections between wake structure and kinematics are also shown in the motion of oscillating foils mimicking flapping flight of insects, in which the size and position of shed vortices are a function of frequency of oscillation \cite{Wang2000}.
Wake studies specific to higher heave amplitudes in pitching and heaving foils have shown that a strong reverse von K\'arm\'an street is associated with higher thrust generation whereas a distorted wake is associated with degradation in thrust generation \cite{Hover2004,Xiao2010}.

\red{However it has also been demonstrated that there is not always a one-to-one correspondence between thrust and wake structure. Floryan et al. give counterexamples that show large differences in the vortex wake structure with no change in performance, and conversely, that subtle changes can have great changes in performance \cite{Floryan2020}. Also mentioned are other factors that can decouple the wake from accurately predicting the performance including the Reynolds number, and the subsequent interactions of vortices within the wake which are unlikely to highly influence the drag and thrust on the foil. }

More recently, wake structure classification has been augmented by machine learning utilities. Using information solely from a matrix of point measurements downstream of the wake, Colvert et al. \cite{Colvert2018} classify three specific wake patterns with an artificial neural network consisting of four layers.
Vortex wake classification has also been performed by Wang and Hemati \cite{Wang2019} where a k-nearest neighbors model classifies relative velocity time signals measured from a foil-based sensor. Some of the signals are known to correlate with well-known wake structures, however when the classification algorithm is applied to new exotic patterns, it produces high accuracy results.
\red{Li et al. \cite{Li2020} classified the wake patterns from computations of a fish-like flexible swimmer by training a fully connected neural network with one-time point measurements of velocity and vorticity in the wake.}

The use of classification techniques such as the ones described above rely on the user to define the classes of wake patterns and select the appropriate class for specific kinematics. A classification model is subsequently trained with a subset of the data, known as the training set. The trained model is then utilized to sort kinematics from rest of the subset, known as the test set, which provides an estimate for the accuracy of the trained model.
These methods would not be appropriate for discovery of chaotic or hybrid wake modes, for example those that result from multiple degrees-of-freedom or non-sinusoidal kinematics that may not be easily characterized with the $mS + nP$ notation, or may not be discernible by a human. As an example, the vortex wake patterns behind oscillating foils for energy harvesting applications are often generated with high-heave amplitudes and exotic wake patterns \cite{Ribeiro2020}. These challenges are exacerbated due to non-uniform or turbulent inflow conditions at high Reynolds numbers.

In lieu of classification, this paper uses {\it clustering}, which automatically groups similar samples without prior definition of the cluster labels and without supervised training. \red{This method does not rely on the $mS + nP$ notion of counting vortices, nor does it have any pre-determined criteria defined in grouping similar wakes or kinematics together.} The samples are images representing snapshots of wake vorticity, taken from computational fluid dynamics (CFD) simulations of pitching/heaving foils under various kinematic conditions. The clustering is performed with a k-means++ algorithm \cite{arthur2006}, and is supported by a convolutional autoencoder to reduce the the images to a latent space of pertinent features.

\begin{figure}
    \centering
    \begin{subfigure}{0.64\textwidth}
        \centering
        \includegraphics[width=\textwidth]{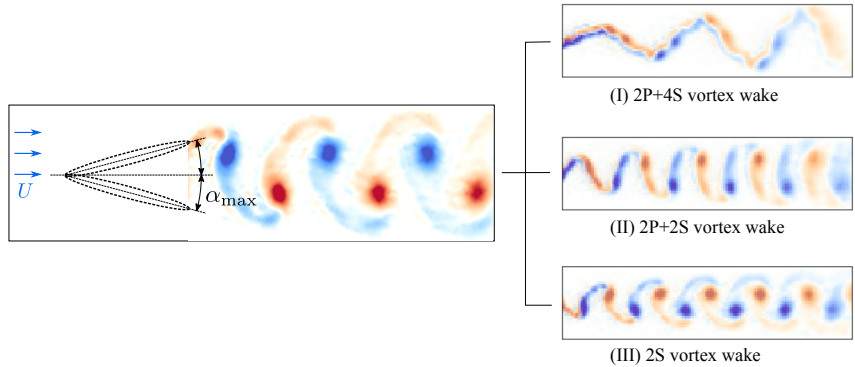}
        \caption{One Degree-of-Freedom - Pitching - Prelabeled Dataset}
        \label{fig:sketch1}
    \end{subfigure}
    \par\bigskip
    \begin{subfigure}{0.55\textwidth}
        \centering
        \includegraphics[width=\textwidth]{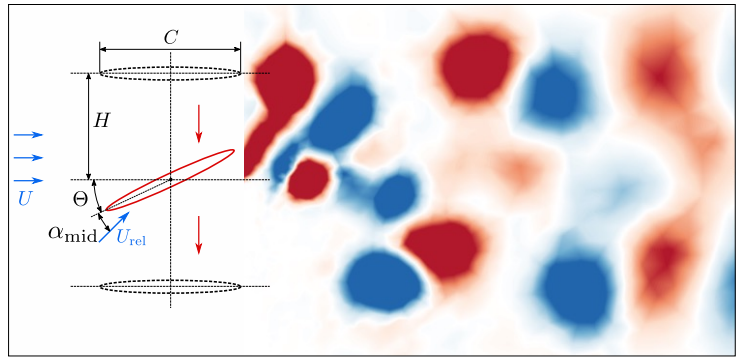}
        \caption{Two Degrees-of-Freedom - Pitching/Heaving - Unlabeled Dataset}
        \label{fig:sketch2}
    \end{subfigure}
    \caption{Oscillating foil propulsion datasets used for the clustering. \textbf{(a) One degree-of-freedom: pitching only.} This dataset generates three prelabeled vortex wake patterns that are a function of frequency ($\alpha_{max}$ remains constant). \textbf{(b) Two degrees-of-freedom: pitching and heaving.} This dataset generates a range of complex wake patterns by varying pitch amplitude, heave amplitude, and frequency.}
    \label{fig:kinematics1}
\end{figure}

Figure \ref{fig:kinematics1} displays the two datasets of wake images that are analyzed. 
In order to tune and test the model, a prelabeled dataset is developed using the same oscillating foil kinematics as Colvert et al. \cite{Colvert2018}. This dataset represents wakes from \red{computations of} a one degree-of-freedom \red{pitching foil} and produces three distinct and well-defined wake patterns as shown in figure \ref{fig:sketch1}.  Subsequently, the \red{tuned} model is applied to vorticity wake images from a CFD database of thrust-generating oscillating foils with two degrees-of-freedom, pitch and heave, as shown in figure \ref{fig:sketch2}. This larger and unlabeled dataset is comprised of more complex vortex wakes (see Dave et al. \cite{Dave2020}) represented in figure \ref{fig:sketch2}. 

Using the autoencoder and clustering pipeline, the wake images are automatically sorted into groups. 
Since wake patterns are correlated with propulsive performance, the resulting cluster placement is expected to determine the performance regime in which the oscillating foil is operating based solely on the wake image. The clustering results are mapped onto a space of dominant kinematic variables and performance metrics (thrust and efficiency), demonstrating an automated sorting of the vortex images into distinct kinematic and propulsive regimes.

The paper is organized to describe the computational methods used to create the datasets, followed by a statistical correlation between the kinematics and performance metrics. The autoencoder and clustering methodology is then described, followed by the clustering results.

\section{Computational Methods}\label{sec:CFDmethods}

\subsection{CFD solver}

Two databases are produced by simulating the flow field around an oscillating foil with CFD. \red{The first is a {\it prelabeled dataset} formulated from a pitching foil with only one degree-of-freedom.  The purpose of the prelabeled dataset is to tune hyper-parameters of the autoencoder and clustering on data with distinct wake structures easily classified with visual inspection. The main focus of the paper is on the second dataset, or the {\it unlabeled dataset}. This data is comprised of flow fields around a pitching and heaving foil with two degrees-of-freedom, generating more complex and chaotic wake structures over a much wider range of kinematic flapping parameters. Once the CAE and clustering algorithm have been developed and tuned using the labeled dataset, the analysis methods are subsequently applied to the larger unlabeled dataset.} 

\begin{figure}[h]
    \centering
    \includegraphics[width=0.64\textwidth]{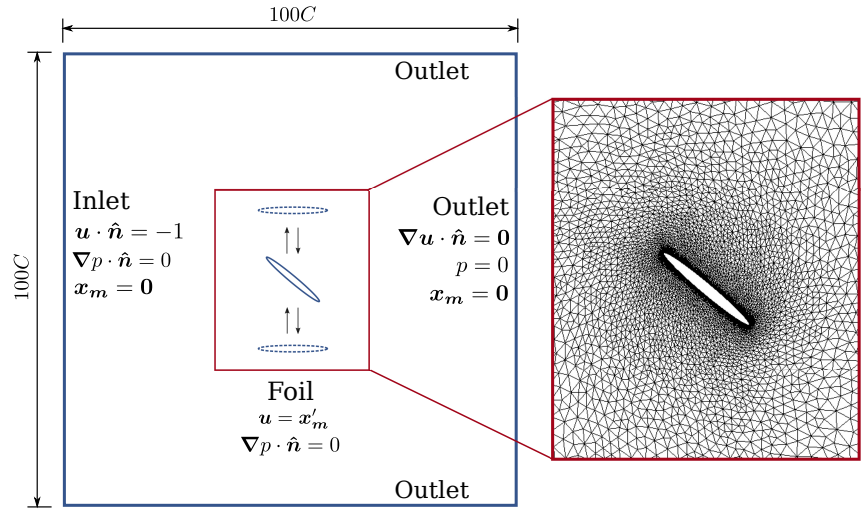}
    \caption{Simulation domain (not to scale) and mesh configuration during the upstroke.}
    \label{fig:CFDdomain}
\end{figure}

Simulations are performed with a second-order accurate open-source finite volume solver from the {\it OpenFOAM} package. To solve for the fluid flow around a moving foil, a dynamic mesh library was utilized that dynamically computes the displacement of each mesh element, $\bm x_m$, at every time step driven by the motion of the oscillating foil.  Kinematics of the foil are translated to a rigid body motion of each element on the surface of the foil, and implemented as a boundary condition. The mesh motion equation is derived from the linear elasticity equation of motion \cite{Dwight2009}, and includes a diffusivity constant that scales inversely with distance from the foil, such that the cells in the immediate vicinity of the body are not dramatically skewed \cite{Bos2010}. 
Figure \ref{fig:CFDdomain} displays the two-dimensional computational domain, which is 100 chord lengths ($100C$) in each coordinate direction, with inlet and outlet boundary conditions as denoted in the figure.  

\subsection{Prelabeled dataset: Pitching foil}

This dataset is obtained by solving the incompressible Navier-Stokes equations of flow over a NACA0012 foil as it pitches about the leading edge. The simulation is performed at a Reynolds number, $Re=UC/\nu=1000$, where $\nu$ is kinematic viscosity of the fluid, $U$ is the freestream velocity, and $C$ is chord length of the foil. 
\red{The specific kinematics for this prelabeled data is chosen to replicate those simulated and previously classified by Colvert et al. \cite{Colvert2018}. The 15 simulations, each at a different kinematics, can be classified by three distinctly identifiable wake patterns using the ``$mS + nP$" notation.}

As shown in figure \ref{fig:sketch1}, the maximum pitching amplitude, $\alpha_{max}$, is held constant at $5.7^\circ$. The kinematics are varied by changing the frequency of oscillation, $F$, which is non-dimensionalized as the reduced frequency, $F^+=FC/U$. Fifteen frequencies are tested, resulting in fifteen unique kinematics that can be sorted into three classes of wake patterns visualized in figure \ref{fig:sketch1} \red{and listed in table \ref{table:prelabeledset}.} The low frequency cases exhibit a $2P + 4S$ mode, the intermediate frequency range has a $2P + 2S$ mode, and the highest frequencies tested exhibit a $2S$ mode, or a reverse von K\'arm\'an vortex street. 

\red{The mesh for the low Reynolds number simulations in this dataset is comprised of 74,728 cells, with 1022 cells defining the foil surface. The mesh size within the boundary layer is 0.002 chord lengths, and is gradually increased radially away from the foil. The mesh resolution of the wake is adequate to clearly resolve three wavelengths of the vortex wake at the lowest flapping frequency.}

\begin{table}
\centering
\caption{The prelabeled dataset contains 15 simulations, 5 from each regime, that were performed by varying frequency of a foil pitching about its leading edge. These cases generate three distinct vortex wake patterns shown in figure \ref{fig:sketch1}.}
\label{table:prelabeledset}
\begin{tabular}{|p{4.5cm} | p{4.5cm}|} 
\hline 
\multicolumn{2}{|c|}{\cellcolor[HTML]{9B9B9B} {\color[HTML]{333333} \textbf{Prelabeled dataset}}} \\ \hline 
\textbf{Wake Pattern} & $\boldsymbol{F^+}$ \textbf{range}  \\  \hline 
2P + 4S & $0.317 - 0.383$  \\ \hline
2P + 2S & $0.525 - 0.558$  \\ \hline
2S & $0.592 - 0.658$  \\ \hline
\multicolumn{2}{|c|}{\cellcolor[HTML]{EFEFEF} \textbf{Simulation Properties: $Re = 1000$; $B/C=0$; $\alpha_{max}=5.7^\circ$}}\\ \hline
\end{tabular}
\end{table}

\subsection{Unlabeled dataset: Pitching and heaving foil}

The second, unlabeled dataset is designed to produce a richer variety of propulsive wake structures. To accomplish this, sinusoidal motion of a 10\% thick ellipse is prescribed in both the pitch and heave directions, with variations in pitch amplitude ($\Theta$), heave amplitude ($H$), and frequency as depicted in figure \ref{fig:sketch2} and outlined in table \ref{table:heavingset}. 
\red{The phase difference between the heave and pitch stroke, $\phi$, is 90 degrees such that the foil is at zero pitch at the top and bottom of its heave stroke and at maximum pitch at mid-stroke.}
Due to the added degree of freedom in the heave direction, the Strouhal number, $St=2 H F/U$, is used as a non-dimensional parameter representing the maximum heave velocity.
A function of both the frequency and heave amplitude, $St$ has been shown to be correlated with propulsive performance in literature \cite{Koochesfahani1989,Lai1999,Ramamurti2001}.
High-heave kinematics ($H/C \ge 1$) are the predominant focus of the unlabeled data, as they are found to add a high degree of variability into the wake structure and are far less studied in previous literature.  However, an additional set of kinematics at lower heave amplitude ($H/C=0.75$) and different pitch axis location, $B/C=0.33$, is also included in the statistical correlation analysis in section \ref{sec:SVR}. 

\begin{table}
\centering
\caption{The unlabeled dataset contains 63 different combinations of pitch amplitude ($\Theta$) and frequency ($St$) for each of two heave amplitudes ($H$) to produce $126$ unique sets of kinematics.} 

\label{table:heavingset}
\begin{tabular}{|p{2.5cm}| p{2.5cm} |p{2.5cm}|} 
\hline 
\multicolumn{3}{|c|}{\cellcolor[HTML]{9B9B9B}{\color[HTML]{333333} \textbf{Unlabeled dataset}}} \\ \hline 
$\boldsymbol{H/C}$ & $\boldsymbol{\Theta}$ \textbf{range}  & $\boldsymbol{St}$ \textbf{range}  \\ \hline 
$1.0$ &  $15^{\circ} - 55^{\circ}$ & $0.2 - 0.6$ \\ \hline
$2.0$ &  $30^{\circ} - 70^{\circ}$ & $0.4 - 1.2$ \\ \hline
\multicolumn{3}{|c|}{\cellcolor[HTML]{EFEFEF}\textbf{Simulation Properties: $Re = 10^6$; $B/C = 0.5$; $\phi=90^{\circ}$}}\\ \hline
\end{tabular}
\end{table}

\begin{figure}[]
  \centering
  \includegraphics[width=0.4\textwidth]{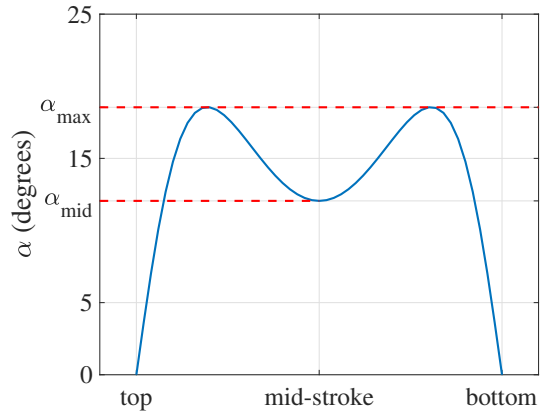}
  \caption{Variation in the relative angle of attack during a downward stroke of the foil for $H/C = 1,\ St = 0.6,\ \Theta=50^\circ$.}
    \label{fig:alpha_rel}
\end{figure}

Due to the changing relative velocity vector, the relative angle of attack of the flow with respect to the foil, $\alpha$, varies over the cycle as demonstrated by figure \ref{fig:alpha_rel}. As noted in previous work \cite{Dave2020,Hover2004,Xiao2010}, the relative angle of attack can be used as an indicator of propulsive performance. Over the span of the high heave and high pitch kinematics tested, the shape of $\alpha$ throughout the cycle can take various forms, and thus two critical values are reported for each set of kinematics: the maximum value, $\alpha_{max}$, and value at mid-downstroke, $\alpha_{mid}$.

An unstructured mesh is generated with approximately $80,000$ cells, \red{with a high density} clustered close to the foil to ensure adequate boundary layer resolution. To demonstrate the mesh structure and resolution in close vicinity of the foil, figure \ref{fig:CFDdomain} illustrates the foil position and deformed mesh configuration during upstroke. \red{The maximum value of non-dimensional wall distance, $y^+$, at the first mesh layer on the foil is 1.87 when the foil is stationary at zero angle of attack, and the number of circumferential mesh cells along the foil boundary is 6774.
A mesh resolution study has been previously performed \cite{Dave2020} where it is demonstrated that the unsteady forces and thrust generation are adequately resolved at the current resolution. Furthermore, the wake vorticity is compared between the current mesh and one with higher resolution in the far-field. The comparison highlights that the same vortex structure persists, although the strength of the vortices is weakened with the slightly lower mesh resolution. This has a similar effect as varying the RANS turbulence model which is discussed in more detail below. The time step for the computation is determined dynamically by imposing a maximum Courant number of 0.95.}

\begin{figure}[]
  \centering
  \includegraphics[width=0.72\textwidth]{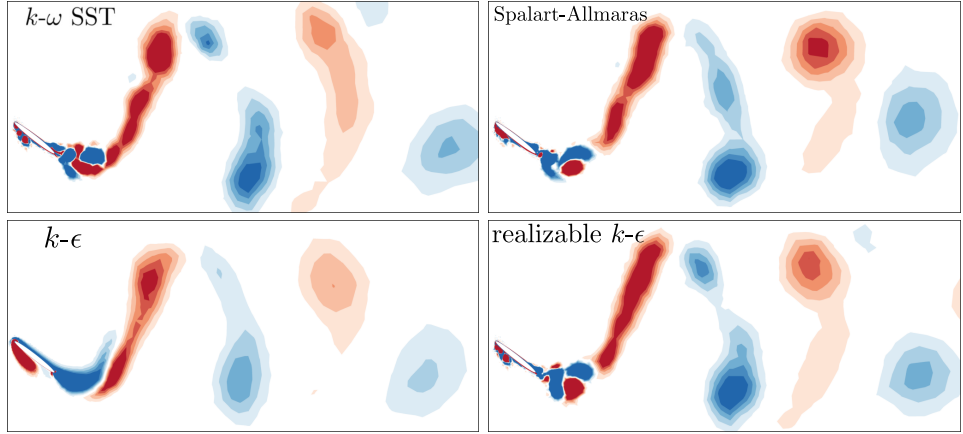}
  \caption{\red{Instantaneous spanwise vorticity ($\Omega C/U$) fields from computations with four different turbulence models for the kinematics $H/C = 1,\ St = 0.6,\ \Theta=45^\circ$. Contours range from -2 (blue) to 2 (red).} }
    \label{fig:models_wake}
\end{figure}

This dataset is performed at a high, turbulent Reynolds number of $10^6$ using a Reynolds-averaged Navier-Stokes (RANS) solver with $k$-$\omega$ SST turbulence closure.
\red{The $k$-$\epsilon$, realizable $k$-$\epsilon$, and Spalart Allmaras models were also tested for a single set of kinematics to examine model sensitivity. The $k$-$\epsilon$ model was too dissipative and the thrust computed, $C_T=1.22$, deviated significantly from the other three models, which were very close to each other with $C_T\approx1.42$. Figure \ref{fig:models_wake} compares instantaneous vorticity fields from computations using the four models. Except for $k$-$\epsilon$, there are only minor differences among the remaining three models in terms of the vortex formation during oscillation, and the overall structure and location of vortices aligns well. It is expected that these small differences would be further diminished once the images are processed by the autoencoder and reduced to a latent space for clustering.}
More details on this dataset including the simulation set-up, validation, mesh refinement, and comparison against experimental data, are outlined in Dave et al. \cite{Dave2020}.

\subsection{Performance metrics}

The kinematics chosen for the unlabeled dataset fall within the thrust-producing regime, in which the net propulsive force created by the foil's motion can be characterized by the thrust coefficient, 

\begin{equation}
      C_T = \frac{<F_x>}{\frac{1}{2} \rho U^2 C},
\end{equation}

\noindent where $<F_x>$ is the time-averaged force in the horizontal direction, opposing the freestream flow.  There is power exerted to generate thrust, characterized by the input power coefficient, $C_P$. The power coefficient is computed as the work required to perform the vertical and pitching motions, or

\begin{equation}
    C_P = -\frac{<F_y(t)\frac{dh(t)}{dt} + M_z(t)\frac{d\theta(t)}{dt}>}{\frac{1}{2} \rho U^3 C},
\end{equation}

\noindent where $F_y(t)$ and $M_z(t)$ are the vertical force and the moment during the heave, $h(t)$, and the pitch, $\theta(t)$, motions. The time-averaged input power will be denoted by $C_P$. Furthermore, the amount of input power that is converted to thrust is defined by a propulsive efficiency, 

\begin{equation}
    \eta = \frac{C_T}{C_P}.
\end{equation}
\section{Correlation of Kinematic Parameters with Performance Metrics}\label{sec:SVR}

In order to better correlate parameters that most strongly influence the variation in performance and hence the wake structure, statistical analysis is performed on the unlabeled two degrees-of-freedom dataset. Three independent variables, $\Theta$, $H/C$, and $St$, are varied directly, and the secondary kinematic parameters, $\alpha_{mid}$ and $\alpha_{max}$, are computed.  For performance metrics, the propulsive efficiency, $\eta$, the power input, $C_P$, and thrust coefficient, $C_T$, are all evaluated. 

\begin{figure}[]
    \centering
    \begin{subfigure}[t]{0.36\textwidth}
        \centering
        \includegraphics[width=\textwidth]{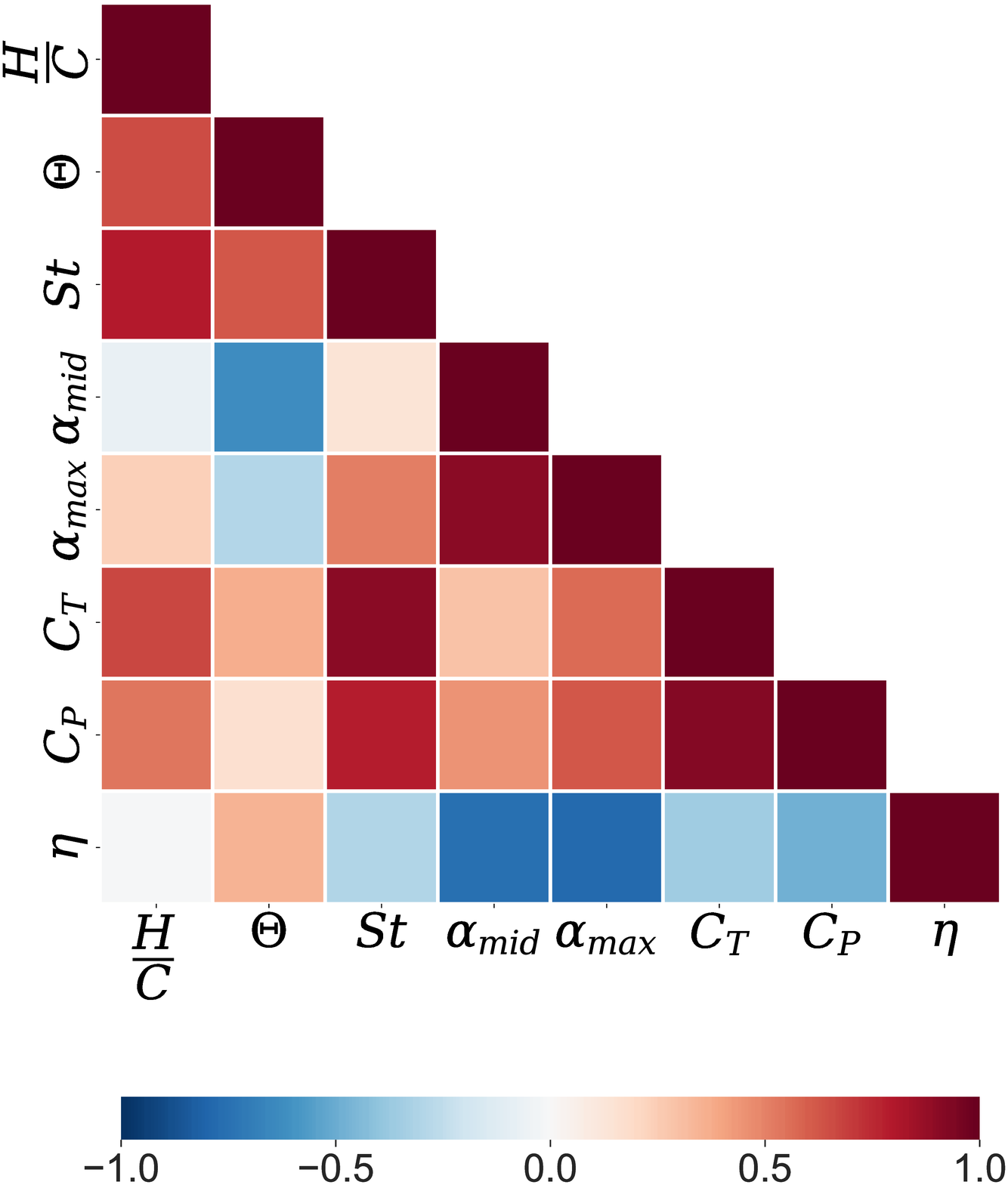}
        \caption{Correlation Matrix}
        \label{fig:correlationsa}
    \end{subfigure}
    \hspace{0.25cm}
    \begin{subfigure}[t]{0.6\textwidth}
        \centering
        \includegraphics[width=\textwidth]{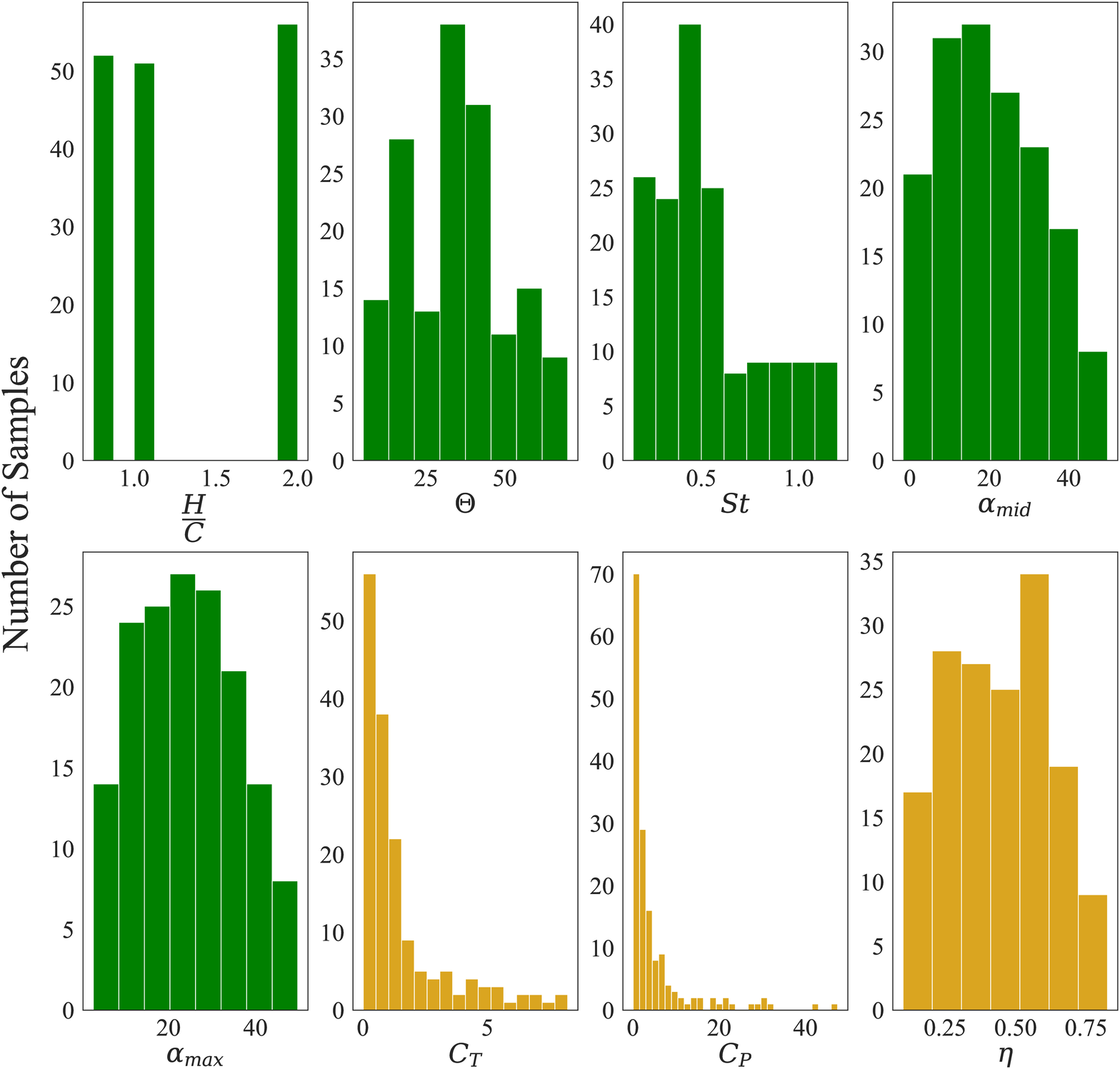}
        \caption{Histograms}
        \label{fig:correlationsb}
    \end{subfigure}
    \caption{(a) Correlation Matrix: Pearson correlation coefficients among the kinematic parameters and output metrics, with a value of $-1$ indicating inverse correlation and $1$ indicating direct correlation. 
    (b) Histograms: Frequency of samples in different bins or range of values for each parameter studied. In green - kinematic parameters, in gold - performance parameters.}
    \label{fig:correlations}
\end{figure}

To discover which kinematic parameters most strongly influence performance, a matrix of Pearson correlation coefficients is computed in figure \ref{fig:correlationsa}. This matrix displays a pairwise statistical correlation 
between two individual variables in terms of magnitude and direction of the association. Values close to $1$ are directly correlated, values close to $-1$ are inversely correlated, whereas values close to $0$ do not show a correlation at all.

\red{As a result of this matrix, the inverse correlation between $C_T$ and $\eta$ is observed here as is also well established in literature. Additionally, it is seen that the most influential parameter to increase $C_T$ is an increase in $St$, 
which is consistent with other findings e.g. \cite{Ramamurti2001,Schouveiler2005,Techet2007}. The input power, $C_P$, is also strongly dependent on $St$, but not to the same degree as the thrust coefficient. 
The increase of thrust and input power with $St$ has been explained using scaling relations by Floryan et al. \cite{Floryan2019}. This work shows an increase in $\eta$ with respect to amplitude at high $St$, but the opposite trend with low $St$, explained by the influence of drag. These scaling laws are relevant only for lower relative angles of attack, or kinematics that do not display large flow separation.}

\red{In contrast, the current research explores a frequency range and heave amplitude range that commonly result in high relative angles of attack inducing massive flow separation. For the kinematics explored the pitch amplitude has a slightly positive correlation with $\eta$ however there is no correlation between heave amplitude and $\eta$. 
Hence it is observed that the thrust coefficient increases with the heave amplitude at no cost to the propulsive efficiency in an average sense. To independently establish the role of heave compared with pitch, a more systematic approach would have to be employed that holds $St$ constant while increasing heave over a wider range of amplitudes.
}
The relative angle of attack is also important in prediction of the output kinematics, as shown by many others in the literature e.g. \cite{Schouveiler2005}. Figure \ref{fig:correlationsa} confirms these findings by demonstrating that a decrease in either $\alpha_{max}$ or $\alpha_{mid}$ has the strongest impact on increasing $\eta$. In terms of deciphering the effects of $\alpha_{max}$ versus $\alpha_{mid}$, the correlation matrix demonstrates the strong inverse relationship of $\alpha_{mid}$ with $\Theta$, whereas $\alpha_{max}$ has a strong proportional relationship to $St$ for the kinematics range explored.

It is important to note that the statistical correlations of figure \ref{fig:correlationsa} are valid only for the range of parameters available in the unlabeled dataset. Thus, figure \ref{fig:correlationsb} shows the data sample distribution for each of the kinematic parameters in green and the output metrics in gold. The kinematic parameters are fairly well-distributed with the exception of heave amplitude, $H/C$, in which only three distinct values are explored. The output metrics $C_T$ and $C_P$ are more heavily concentrated at low values, whereas propulsive efficiency has a wider distribution. Including more data at lower heave amplitudes may skew the parameters and correlations in a slightly new direction, however this paper specifically focuses on the thrust-producing kinematics at high heave amplitudes. The narrow focus on high heave kinematics imposes a tractable limit to the number of simulations while still generating a variety of wake modes for the clustering. 
\section{Autoencoder and Clustering Methodology}\label{sec:clustering}

\subsection{Data pre-processing}

\red{The input into the convolutional autoencoder are images of the spanwise vorticity fields of the wake behind the oscillating foil. Vorticity is plotted in the \textit{Tecplot} software, with 5 contours from blue (RGB 0, 0, 255) to red (RGB 255, 0, 0). In the unlabeled dataset the window size, kept constant at 8.5C by 11C, is large enough to accommodate the highest heave cases with the foil stroke centered vertically, and aligns the \red{trailing} edge of the foil on the left border of the frame. The horizontal frame width is large enough to capture at least one wavelength of the resulting wake for the lowest oscillation frequency. A final post-processing script floods the zero-vorticity background to black (RGB 0,0,0) and reduces the resolution to 128 by 128 pixels for the final version of the PNG files used by the autoencoder. }

\red{The prelabeled dataset contains 15 unique kinematics, each sampled with 14 frames, for a total of 210 images. 
The unlabeled dataset samples each unique set of kinematics 21 times over a half stroke. The heave amplitudes of $H/C=1$ and $H/C=2$ are considered in the clustering algorithm 
with 121 unique kinematics, 
resulting in 2541 image samples. It is important to note that each of these image samples is clustered independently and not as a group with the other 20 samples from the same kinematics.}

\subsection{Autoencoder architecture}

The clustering methodology consists of a deep convolutional autoencoder (CAE) \cite{Alqahtani2018} used for image feature extraction followed by a k-means++ algorithm to cluster similar images in the embedded feature space. The convolutional autoencoder described below is implemented using the Python based libraries \textit{TensorFlow} \cite{tensorflow2015-whitepaper} and \textit{Keras} \cite{chollet2015keras}.

An autoencoder falls within the broader category of neural networks, however it often has the specific goal of reducing a complex image to a compact representation of the most dominant features \cite{NIPS2012_4824, Zeiler2014}, which in this case is the fundamental vortex wake structure. 
Pertinent features within the image are detected with filters, which create feature maps through a convolution operation on the preceding layer. Each convolutional layer utilizes multiple filters, hence generating multiple feature maps with the same height and width, and the number of maps defines the depth. Additionally, the max-pooling layers perform a downsampling operation with a $2\times 2$ filter along the width and height dimensions, resulting in a reduction of factor $2$ in the size of each dimension, while the depth remains constant.

The combination of convolutional and max-pooling layers increases the neural network performance through many strategies: reducing the feature space to a more manageable size, reducing the number of parameters in the network to control for overfitting, making the CAE invariant to small distortions and translations in the input image, and helping the CAE to arrive at an almost scale invariant representation of the input images (detecting features irrespective of their size in the image).

\begin{figure}[h]
    \centering
    \includegraphics[width=1\textwidth]{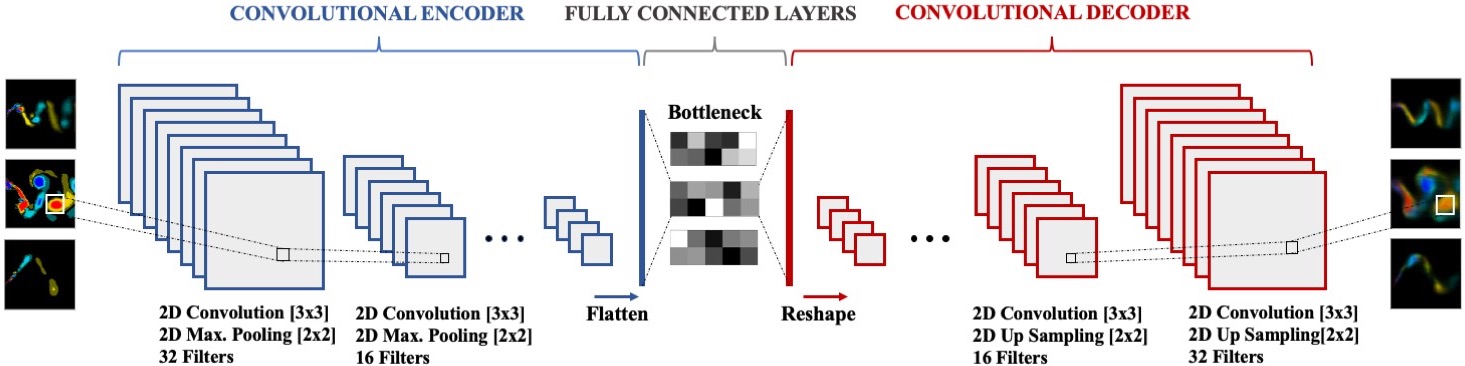}
    \caption{Convolutional Autoencoder architecture.}
    \label{fig:autoencoder}
\end{figure}

Figure \ref{fig:autoencoder} displays a summary of the architecture implemented. The first portion is the convolutional encoder, which consists of five sequences of convolutional and max-pooling layers that compress the original $128 \times 128$ pixel image with 3 color channels
to a $4 \times 4$ feature map with 4 channels. 
The next step is to flatten the 3-dimensional matrix, followed by a fully-connected neural network layer to 
yield a $10$ dimensional feature space, commonly known as the bottleneck. 
The convolutional decoder, which consists of layers in reverse order with exactly the same hyperparameters as used in the encoder, reconstructs the image. Finally, outputs are compared pixel to pixel with the original image inputs using 
a cross-entropy loss function \cite{Nasr2002,Zhang2018}. 

The \textit{Adam} optimization algorithm \cite{Kingma2015} is utilized to iteratively compute the parameter values for the convolution filters as well as the fully connected layers, so as to decrease the loss function and mimic the input matrix as accurately as possible. The number of layers, the filter size, the number of filters in each layer, normalization functions, and use of regularization methods have also been tuned to 
reduce the loss function while increasing the bottleneck compression.
Since the input image is compared with the reconstructed image to reduce the loss function and extract the dominant features at the bottleneck, this step is considered unsupervised.
\ref{Appendix} shows a detailed scheme of the final CAE architecture.

Two indicators of a good autoencoder are a high compression from the input image to the bottleneck and an accurate reconstruction of the original image. A high compression means a low dimensional feature space as an input to the clustering algorithm, whereas a good reconstruction ensures that those features represent enough of the original image. 
The balance between high compression and proper reconstruction is not easy to achieve, especially when the original images have high resolution data resulting in a deeper neural network in order to achieve a high compression. In the current CAE, a strategy based on skip-connections proposed by Xiao-Jiao Mao et al. \cite{Mao2016} was implemented to address this challenge. Skip-layer connections link convolutional and their respective deconvolutional layers, increasing both training speed and reconstruction accuracy. This method allows the loss function signal to back-propagate to bottom layers, directly reducing the problem of vanishing gradients \cite{Hochreiter1998} while keeping a high compression level. It also enables the encoding layers to pass information to the decoding layers, hence increasing the image reconstruction accuracy.

\subsection{k-means ++ clustering}
\label{sec:clustermethods}

Once the data is properly compressed into a low dimensional latent space of features, the resulting arrays are automatically grouped using a k-means++ clustering algorithm. This method was chosen after comparing with results from more sophisticated methods including K-medoids \cite{Jin2010}, hierarchical clustering \cite{7100308}, and DBSCAN \cite{Ester1996}, none of which showed any improvement over k-means++. The clustering algorithm is implemented using the \textit{scikit-learn} library \cite{Pedregosa2011}.

k-means \cite{Lloyd1982} is an iterative method that requires user input to determine $N$, the number of cluster groups. The algorithm subsequently initializes $N$ random centers, which are points in the multidimensional feature space.  
For each data point (in the compressed latent space), the distance to each center is calculated and the point is classified into a group based on minimum distance.
Based on the classified points, the cluster center is recalculated with each iteration by taking the mean of each dimension of all the vectors in the group. k-means++ is an enhanced version of the original k-means algorithm which improves the initial choice of centers, increasing both accuracy as well as speed \cite{arthur2006}.

Thus, the number of clusters, $N$, is predetermined by the user. The preferred number of clusters will be the $N$ value that has a minimum mean intra-class distance and a maximum mean inter-class distance, which are not always concurrent with one another.
The elbow method \cite{Thorndike1953} computes the total within-cluster sum of square (WSS) error, known as distortion, and plots it against $N$, choosing the value of $N$ that is on the `elbow' of the curve as the indicator of the approximate number of clusters. The average silhouette score method \cite{Rousseeuw1987} is often used in tandem with the elbow method to measure the quality of a clustering. This method determines how well each point lies within its cluster by estimating {\it cohesion} and {\it separation} as Euclidean or Manhattan distances. For an individual point, cohesion is the average distance of that point to all the other points in its cluster, and separation is the minimum of its average distance to all the points in another cluster. The Silhouette score is a combination of both factors, typically ranging from $0$ to $1$, where a higher score indicates better clustering. 

\section{Results and discussion}\label{sec:results}
\subsection{Clustering accuracy of prelabeled dataset}
\label{prelabeled}
\begin{figure}[h]
    \centering
    \includegraphics[width=0.9\textwidth]{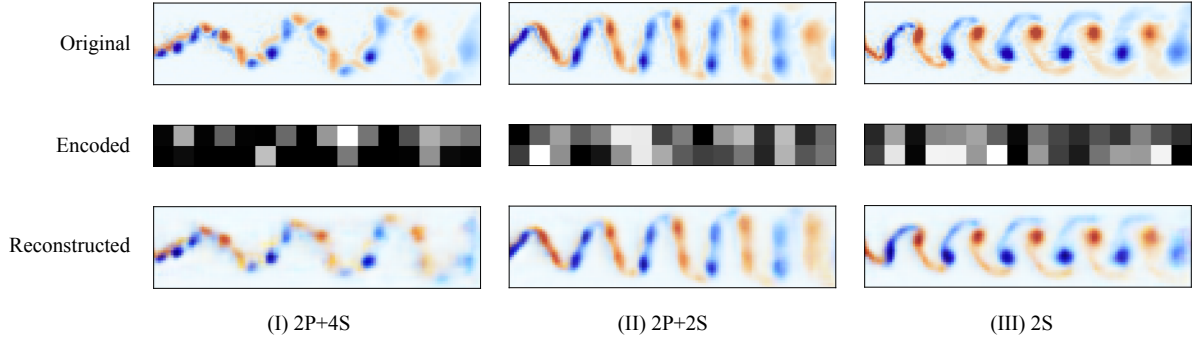}
    \caption{Three different vortex wake patterns for the one degree-of-freedom samples, presorted into categories I, II, and III. Representation of the original images (first row), encoded images or latent space of features (second row), and reconstructed images (third row).}
    \label{fig:encodingSimple}
\end{figure}

The one degree-of-freedom dataset described in figure \ref{fig:sketch1} is comprised of a foil in a pure pitching mode with varying non-dimensional frequency, $F^+$. The resulting database has 15 different kinematics producing wakes that have been prelabeled into three distinct groups based on the vortex structure \cite{Colvert2018}, \red{with 14 images of the wake sampled for each frequency providing a total of 210 images.}  The first row in figure \ref{fig:encodingSimple} demonstrates vortex patterns typical of each clustering group, and the input image into the CAE. 

The clustering is performed on the encoded latent space of features produced by the CAE algorithm, and demonstrated in the second row of figure \ref{fig:encodingSimple}. The encoded arrays shown are each significantly different from one another, each representing a different wake pattern and thus providing the most important features for the k-means++ clustering algorithm. 
The third row of figure \ref{fig:encodingSimple} depicts the reconstructed image and demonstrates that the tuned CAE is able to achieve both a high compression level in the encoding and an accurate image reconstruction. 

The CAE was trained with $138$ samples from this dataset, followed by $42$ samples for tuning the hyperparameters and $30$ samples for validation. The $210$ images were then compressed and the resultant latent space of features was used as input for the k-means++ clustering. In this prelabeled dataset, the number of clusters is known and was set to $N=3$. 

\begin{figure}[t]
    \centering
    \includegraphics[width=0.27\textwidth]{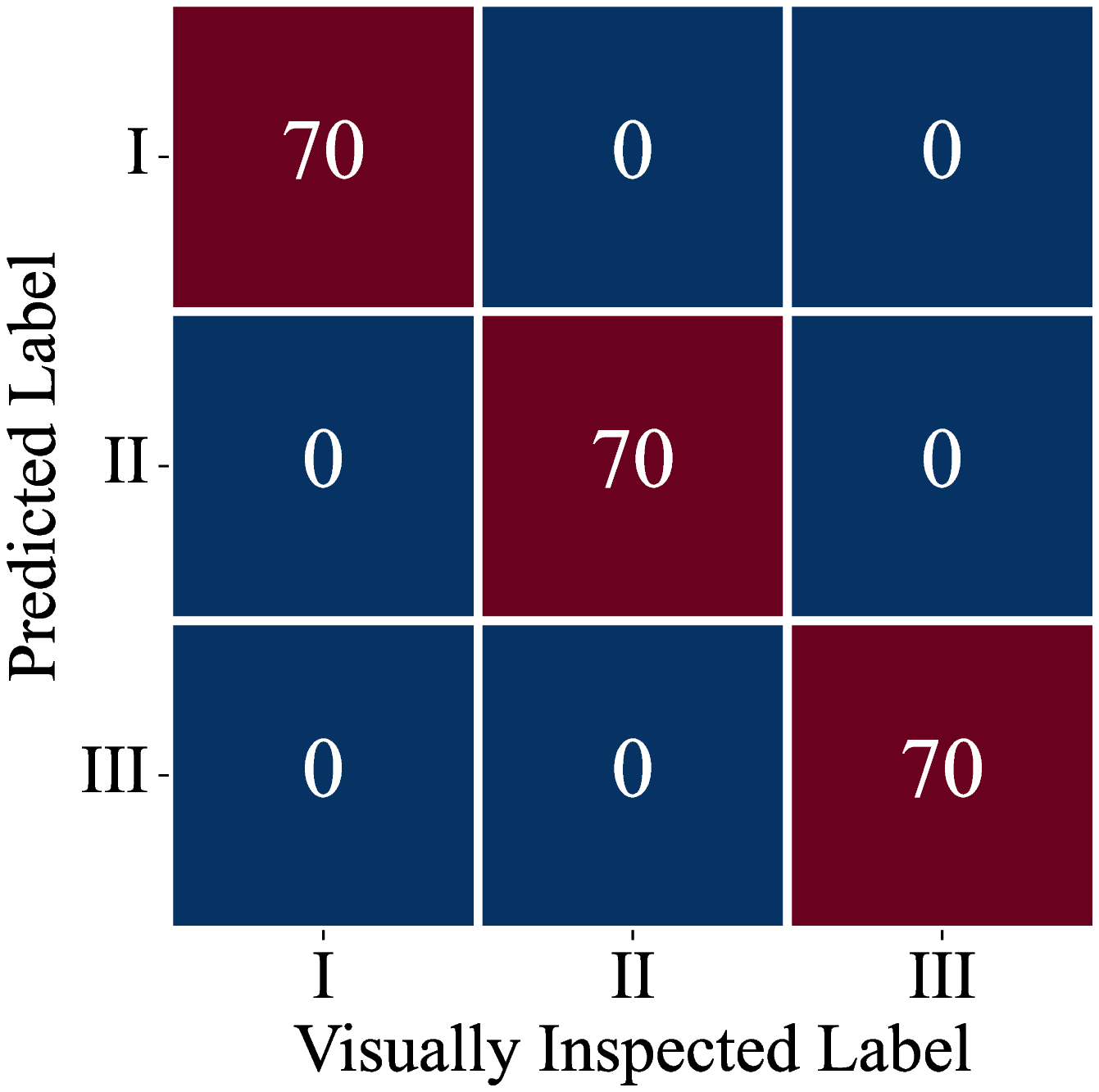}
    \caption{Confusion matrix for the prelabeled one degree-of-freedom dataset.}
    \label{fig:conf1}
\end{figure}

The clustering algorithm grouped exactly $70$ samples in each of the $3$ clusters. All the predicted groupings were coincident with the visually inspected data implying a $100\%$ accuracy for this small dataset, summarized by the confusion matrix in figure \ref{fig:conf1}.

\subsection{Clustering details of unlabeled dataset}

The two degrees-of-freedom model introduced in figure \ref{fig:sketch2} produces a more complex wake structure due to the combined heaving and pitching motion and the higher Reynolds number, resulting in $121$ unique kinematics and a total of 2541 samples of vorticity wake images, that are clustered with the k-means++ algorithm. 
Two different strategies are implemented to analyze the clustering performance as described in the sections below. 

\subsubsection{Validation using visually inspected clustering with $N=4$}

All of the $121$ kinematics were visually labeled and placed in four different groups based on similar vortex wake patterns. A subset of $504$ image samples from the total of $2541$ samples are subsequently labeled. The samples are carefully selected such that there are samples from each of the unique kinematics, and that there are equal number of samples from each group.
 
\begin{figure}
    \centering
    \begin{subfigure}{0.36\textwidth}
        \centering
        \includegraphics[width=\textwidth]{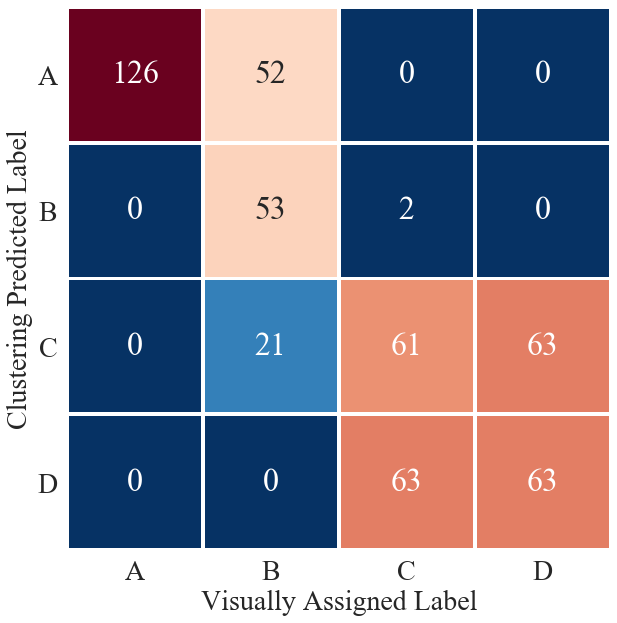}
        \caption{Confusion matrix.} 
        \label{fig:confusionFirstClusteringa}
    \end{subfigure}
    \hspace{1cm}
    \begin{subfigure}{0.45\textwidth}
        \centering
        \includegraphics[width=\textwidth]{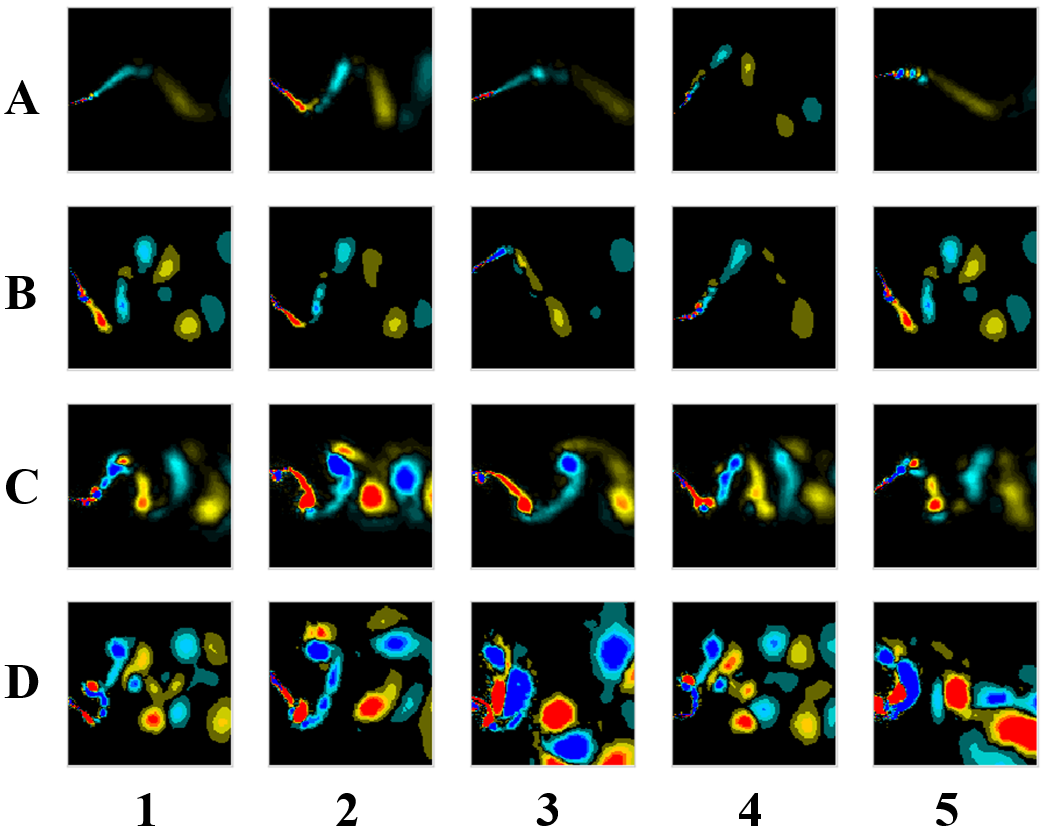}
        \caption{Clusters random samples.} 
        \label{fig:confusionFirstClusteringb}
    \end{subfigure}
    \caption{Results for clustering with $N=4$. (a) Confusion matrix. Predicted clustering compared with visually assigned labels. (b) Random samples from each predicted cluster.}
    \label{fig:confusionFirstClustering}
\end{figure}

The CAE is trained with the remaining $2037$ samples and subsequently used to cluster the $504$ visually labeled samples. The cluster prediction is compared against the visually inspected labels through the confusion matrix shown in figure \ref{fig:confusionFirstClusteringa}. 

The confusion matrix demonstrates that all the samples visually labeled as cluster A coincide with the clustering predictions. However, the results for the other $3$ predicted clusters differ from the visually determined labels. The samples in cluster B are sorted into clusters A, B and C, and the samples from clusters C and D are also intertwined with each other. Some examples of the clustering results are shown in figure \ref{fig:confusionFirstClusteringb} which consists of $5$ random samples from each of the four predicted clusters. 
The samples C4 and C5, for instance, do not exhibit a harmonic vortex wake pattern similar to those in C1, C2, and C3.
However, due to the increased vorticity, the algorithm finds more similarities with the samples in cluster C than the samples in cluster A or B which do not have a clear wake structure.

Even though figure \ref{fig:confusionFirstClusteringa} shows that the predicted labels do not match the visually assigned labels, figure \ref{fig:confusionFirstClusteringb} demonstrates that most of the samples grouped together are quite similar. The discrepancies between predicted and visually determined labels seem to be a consequence of the criteria for sorting (i.e. visual sorting focused on wake structures whereas the CAE and the clustering algorithm placed a stronger value on the wake height and amount of vorticity) rather than a bad algorithm performance.
It also indicates that a different number of wake pattern groups may yield a clearer clustering.

\subsubsection{Clustering with $N=6$}

\begin{figure}
    \centering
    \begin{subfigure}{0.45\textwidth}
        \centering
        \includegraphics[width=\textwidth]{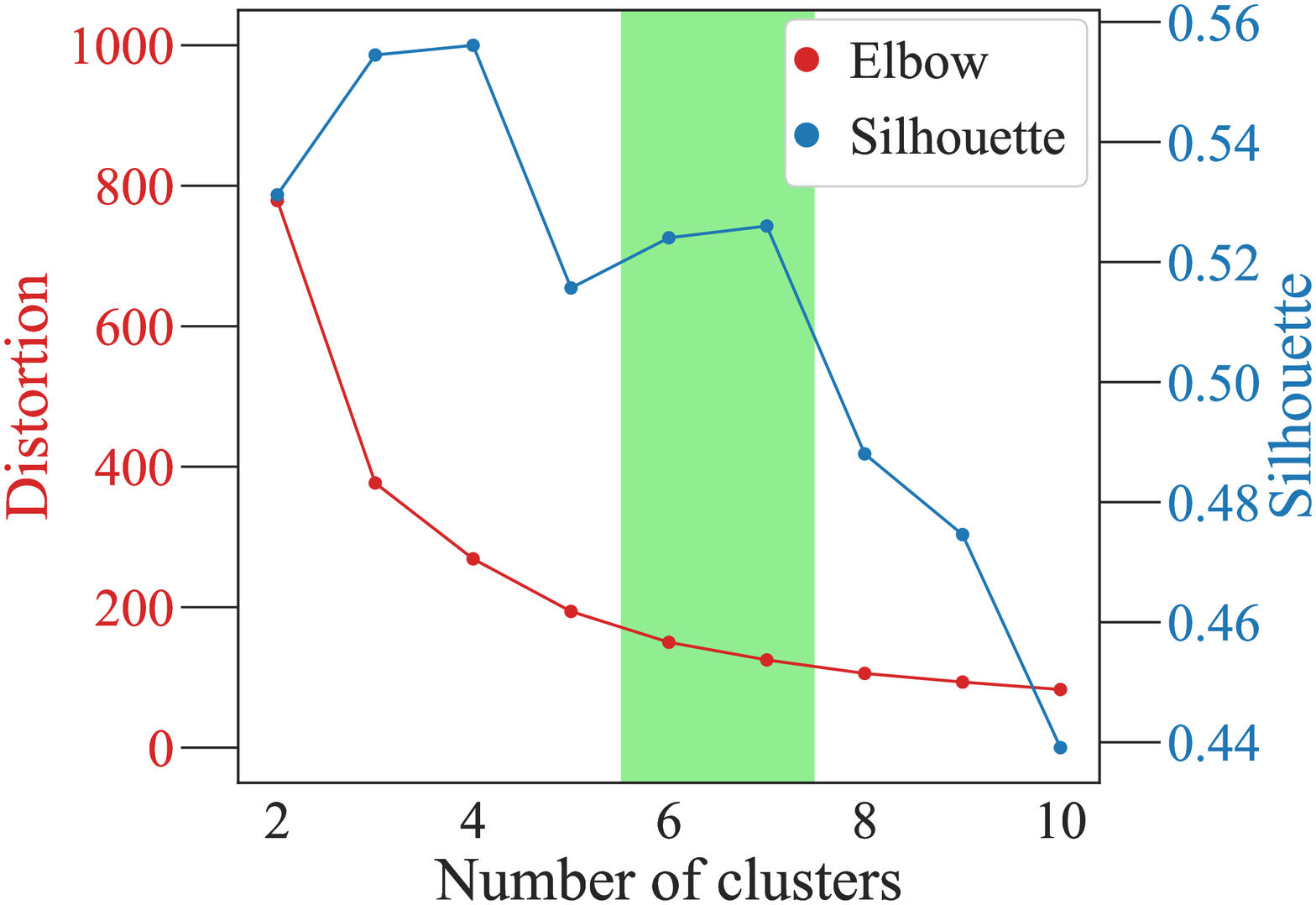}
        \caption{Validation of consistency within clusters.}
        \label{fig:Silhouettea}
    \end{subfigure}
    \hspace{1cm}
    \begin{subfigure}{0.42\textwidth}
        \centering
        \includegraphics[width=\textwidth]{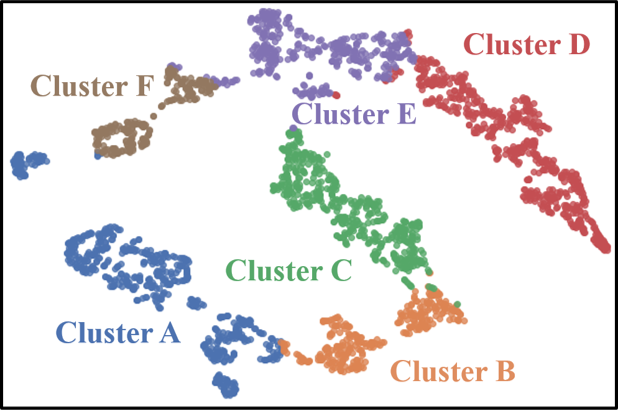}
        \caption{T-distributed Stochastic Neighbor Embedding (T-SNE) of 6 clusters.}
        \label{fig:Silhouetteb}
    \end{subfigure}
    \caption{The optimal number of clusters is determined by a combination of the silhouette and elbow methods, shown in \textbf{(a)}. The optimal region is shaded green indicating $N=6-7$ being the optimal number of clusters for this dataset. The T-distributed stochastic neighbor embedding (T-SNE) distribution is shown in \textbf{(b)} for the 6 clusters.
    }
    \label{fig:Silhouette}
\end{figure}

As a result of the discrepancies between visually assigned and predicted labels with $N=4$, the number of clusters was increased to account for some of the differences unable to be distinguished by only four bins. 
Two techniques introduced in section \ref{sec:clustermethods}, the elbow and silhouette methods, were used together to determine the optimum number of clusters for the dataset. 
Figure \ref{fig:Silhouettea} plots the distortion and silhouette values versus the number of clusters, and the light green region highlights the area that both minimizes distortion and finds a local peak in the silhouette value, indicating that $N=6-7$ is the optimal number of clusters. 
In order to visually display the distribution of the latent space, a T-distributed stochastic neighbor embedding (T-SNE) is applied to the resultant latent spaces in order to further reduce the dimensions into a bi-dimensional embedding. Figure \ref{fig:Silhouetteb} displays the bi-dimensional embedding of the $2541$ latent spaces of features encoded by the convolutional autoencoder, showing the samples which were grouped in each of the $6$ different clusters. The number of clusters was set to $6$ for simplicity since $N=7$ did not significantly improve the final results.

\begin{figure}[bhtp]
    \centering
    \includegraphics[width=0.96\textwidth]{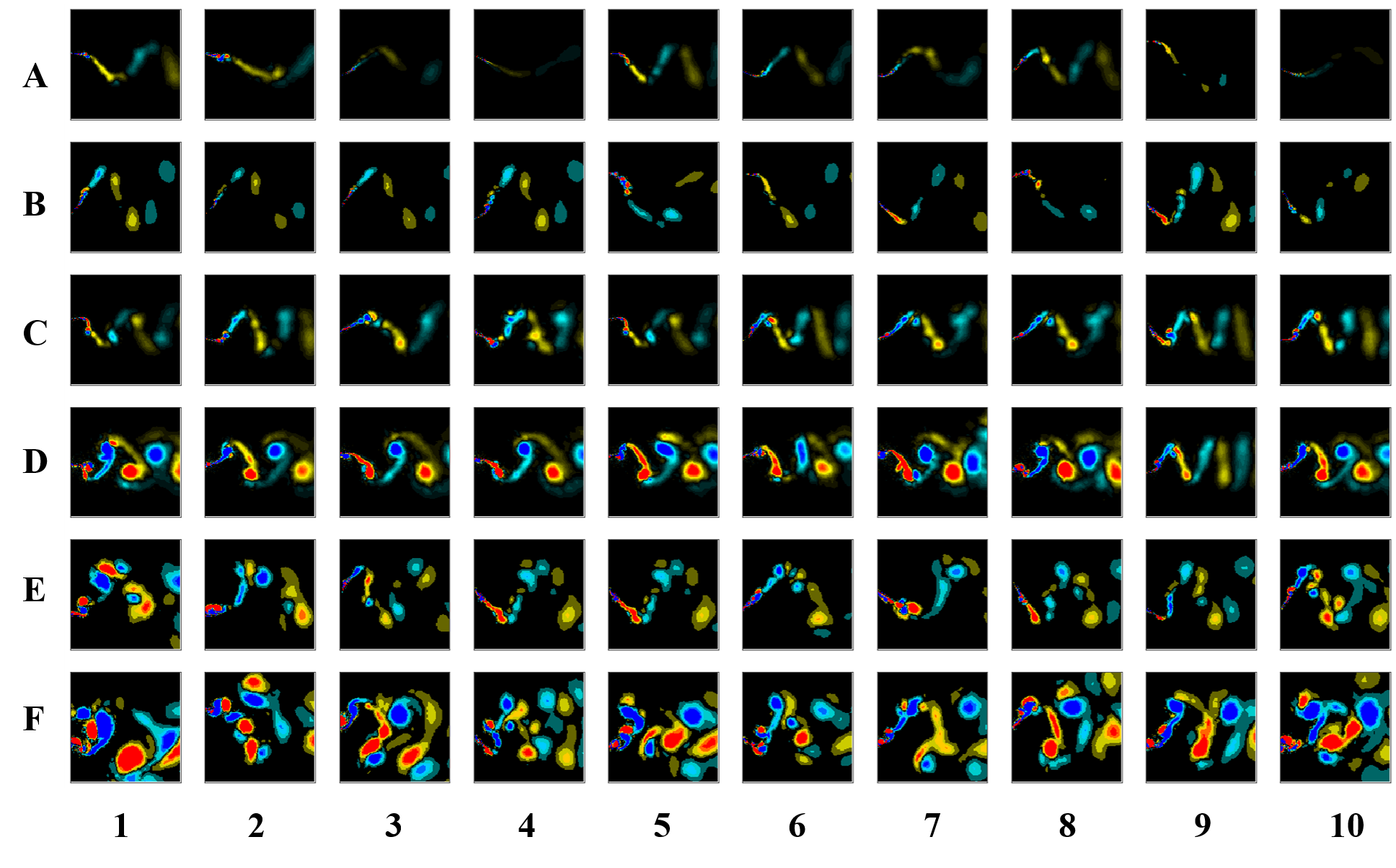} 
    \caption{
    Each row shows 10 random samples from a particular cluster. Most samples clustered together show a similar pattern, though there are exceptions such as sample 9 in Cluster D, which is more similar to the patterns shown in cluster C.}
    \label{fig:10samples}
\end{figure}

Figure \ref{fig:10samples} displays 10 random samples from each predicted cluster. The vortex wakes that are grouped together in the same cluster all exhibit a similar pattern. Clusters A and B group samples with low vortex wake intensity. 
Cluster B differs from cluster A in that the heave, $H/C$, is greater resulting in a stronger vorticity signal in the samples.  
 
Clusters C and D both group samples with more dramatic flow separation than the previous clusters, and both with $H/C=1$. 
The high oscillating frequency also decreases the wavelength between vortices, making these samples easy to distinguish.
For cluster D, vortices in the wake are stronger and the wake pattern is more coherent relative to cluster C. A clear reverse von K\'arm\'an street is also observed likely due to a smoother variation of $\alpha$, in contrast with cluster C that has a more chaotic wake due to a distorted profile of $\alpha$ as shown in figure \ref{fig:alpha_rel}. One exception is sample D9 which is similar to those in cluster C, and may represent a transition between the two clusters. Finally, clusters E and F show chaotic vortex wake structures at the larger heave amplitude, where there is not a clear pattern defined.
Cluster F contains samples with higher vorticity and higher oscillation frequency (as observed from wavelength of the wake).

\subsection{Correlation of clustering with kinematics and propulsive performance}

\begin{figure}[hbtp]
    \centering
    \begin{subfigure}{0.42\textwidth}
        \centering
        \includegraphics[width=\textwidth]{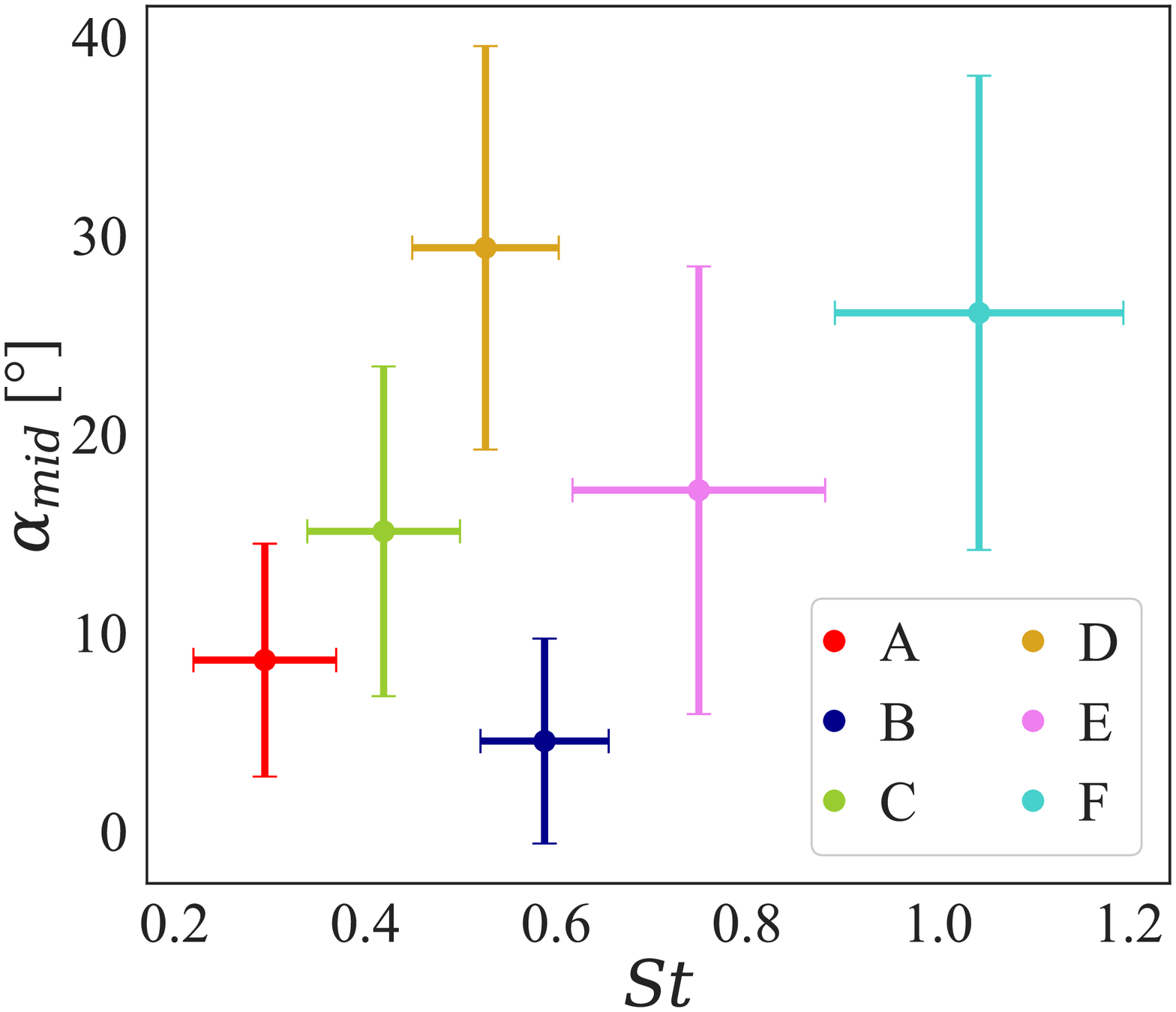}
        \caption{Kinematics space.}
        \label{fig:kinematics}
    \end{subfigure}
    \hspace{1cm}
    \begin{subfigure}{0.42\textwidth}
        \centering
        \includegraphics[width=\textwidth]{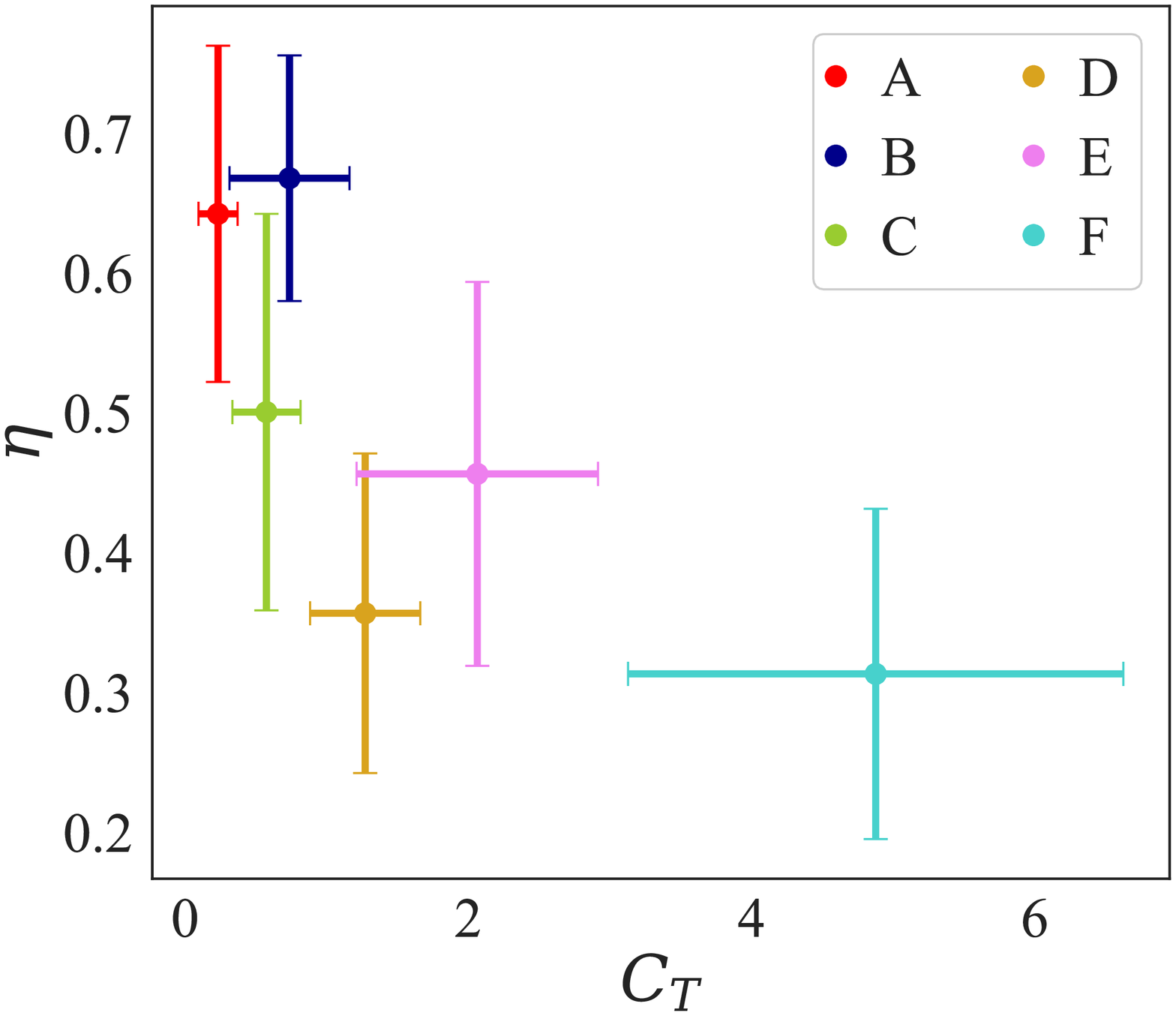}
        \caption{Performance space.}
        \label{fig:performance}
    \end{subfigure}
    \caption{ Map of the clusters in the kinematic and performance spaces. The mean and standard deviation of each parameter are separately calculated for each cluster and presented as an error bar plot, demonstrating how the clusters correspond to different regions in both the kinematic and performance spaces.}
    \label{fig:parameterSpace}
\end{figure}

A method to evaluate the effectiveness of the clustering is to investigate its correlation with the propulsive performance, as it has been observed that propulsive thrust generation is strongly correlated with vortex wake structure \cite{Hover2004,Xiao2010}. In figure \ref{fig:parameterSpace}, each cluster is mapped onto the kinematics space in terms of $St$ and $\alpha_{mid}$, and the performance space in terms of $C_T$ and $\eta$. 
In the kinematics space, each cluster corresponds to a rather unique region with little overlap. 
 
Essentially the clusters A, C, and D group the kinematics with $H/C=1$ in order of increasing frequency (also demonstrated by the increase in $\alpha_{mid}$) whereas clusters B, E, and F do the same for kinematics with $H/C=2$. The wake height is hence the most prominent feature used for clustering the samples, which are then sorted into three groups based on the spread of vorticity and the pattern in the wake. 
Overlaps in regions of performance space are most prevalent between clusters B and C, and another region between D and E. However if both heave amplitudes are observed separately, there is minimal overlap among the clusters A-C-D and among B-E-F.

Clusters A and B are a region with low $\alpha_{mid}$ and relatively low $St$ in the kinematics space. The separation between the two clusters in both the kinematics and performance space is strong, with the primary separation likely due to the two distinct heave amplitudes. The wake patterns are characteristic of the low $\alpha_{mid}$ values indicating that it remains below the static stall angle of attack and there is minimal to no boundary layer separation. 

Both clusters A and B show low $C_T$ and high $\eta$ in the performance space, which is in agreement with the correlations shown in figure \ref{fig:correlationsa} where it was discussed that $\alpha_{mid}$ and $St$ are directly correlated with $C_T$ but inversely correlated with $\eta$. 
Cluster C has a higher $\alpha_{mid}$ than both A or B, and a $St$ range in between the A and B. This causes less than ideal performance in which the efficiency drops significantly and the thrust coefficient is between clusters A and B. Cluster D is similarly characterized, although it is uniquely placed in the kinematics space, its performance has low efficiency and low non-dimensional thrust, and shares some overlap with cluster E in terms of performance.
Finally, clusters E and F, which have the highest $St$ are correlated with the highest thrust generation. Cluster E offers an intermediate range of both efficiency and thrust, with the mean values of $\eta=0.46$ and $C_T=2.06$.  The highest thrust is generated by cluster F, with mean $C_T=4.88$ however with a mean efficiency of $\eta=0.31$.

\section{Discussion and Conclusions}

The results have shown that the pertinent information with regards to performance and kinematics of an oscillating foil can be correlated with the wake structures through machine learning. This method is demonstrated through a CFD database that includes 121 unique kinematic pitching and heaving strokes that span a range of propulsive regimes at $Re=10^6$.  This dataset specifically targets high-pitch and high-heave modes that produce chaotic vortex wakes, further confounded by the high Reynolds number and thus faster rate of vortex break-up when compared with lower Reynolds number flows.  

The methodology utilizes a convolutional autoencoder to distill the information from the wake images into a reduced or latent space.  Using Lasso regularization to avoid overfitting and skip connections to solve the vanishing gradient problem, the autoencoder provides a compressed array composed of $10$ values which is accurately representative of the original images. The images are then clustered based on the representation in latent space using a k-means++ algorithm.  

A one degree-of-freedom model of flow over a pitching foil has three well-defined wake classifications, as previously used by Colvert et al. \cite{Colvert2018}. Using this model, the autoencoder is tuned, and the the clustering demonstrates 100\% accuracy even with a low number of samples used for training. 

\red{The methodology is then applied to the two degrees-of-freedom model that generates vortex wakes from flow over a pitching and heaving foil.} In this model neither the wake classifications nor the number of clustering bins is known.  Using the same hyperparameters as the one-degree-of-freedom model, the autoencoder reduces the images to a latent space and the clustering algorithm automatically sorts the images based on their vortex wake profiles. To assess the results of the methodology, the resulting clusters are mapped to the kinematic space, and to the resulting performance space.  These results show that kinematics with similar performance profiles are clustered together, and the propulsive regimes are clustered into similar groups, supporting the conclusions from previous works that the wake can be a predictive measure of underlying kinematics and propulsive performance. 

The ability to automatically detect subtle differences in the wake patterns of propulsors will help to develop blended, or physics-informed machine learning models for sensing and control methodologies, particularly important for cooperation among multiple foils in an array configuration. In this case the kinematics and performance of each individual foil is dependent on the wake structure (and thus the kinematics and performance) of those foils upstream. Rather than relying purely on data-driven methodologies, understanding the various classifications and formations within the wake will lead to a better understanding of the underlying flow phenomena for predictive modeling. Although oscillating foils are utilized in this paper, these concepts can extend to other bluff body fluid flows where the wake characteristics can be important for the physical and mathematical modeling of the flow. 

While the proposed scheme has proven to work well for the oscillating foil datasets, it has also revealed opportunities for improvement. Due to the interest in the chaotic wakes of high heave amplitudes, there is a low density of data for heave amplitudes (see figure \ref{fig:correlationsb}). Thus, it is likely that more importance is given to the heave amplitude in the clustering algorithm, as the width of the wake is highly influenced by this parameter and can be easily deduced by the vortex wake pattern. Including more samples of varying heave amplitudes may force the clustering algorithm to consider the overall vortex pattern independent of the wake width. However the result may be same, as the heave amplitude does play an important role in the Strouhal number and relative angle of attack, both dominant kinematic factors, and thus clustering heavily based on heave amplitude may prove to be a good strategy. Another strategy would be to employ adaptive windows, and thus purposefully focus on vortical patterns and eliminate the response to wake size.

In this process, the autoencoder hyperparameters are tuned for the simple one-degree-of-freedom model, and then successfully applied to the more complex two-degree-of-freedom model without re-tuning. It has not been determined if the tuned autoencoder would work successfully in more diverse vortex wakes (other than pitching and heaving foils), or other flows in general. However the use of convolutional layers and skip-connections adds robustness such that future applications could be easily trained and hyperparameters determined with minimal training data. 

\section{Acknowledgements}
Part of this research was conducted using computational/visualization resources and services at the Center for Computation and Visualization, Brown University. AGC: Methodology, Analysis, Writing - original draft. MD: Simulations, Analysis, Writing - revising/editing. JAF: Conceptualization, Methodology, Writing - revising/editing.

\section*{References}
\bibliography{newref,ml-mendeley}

\newpage
\appendix
\section{Convolutional Autoencoder Architecture}
\label{Appendix}

\begin{figure}[h]
\centering
    \includegraphics[width=0.75\textwidth]{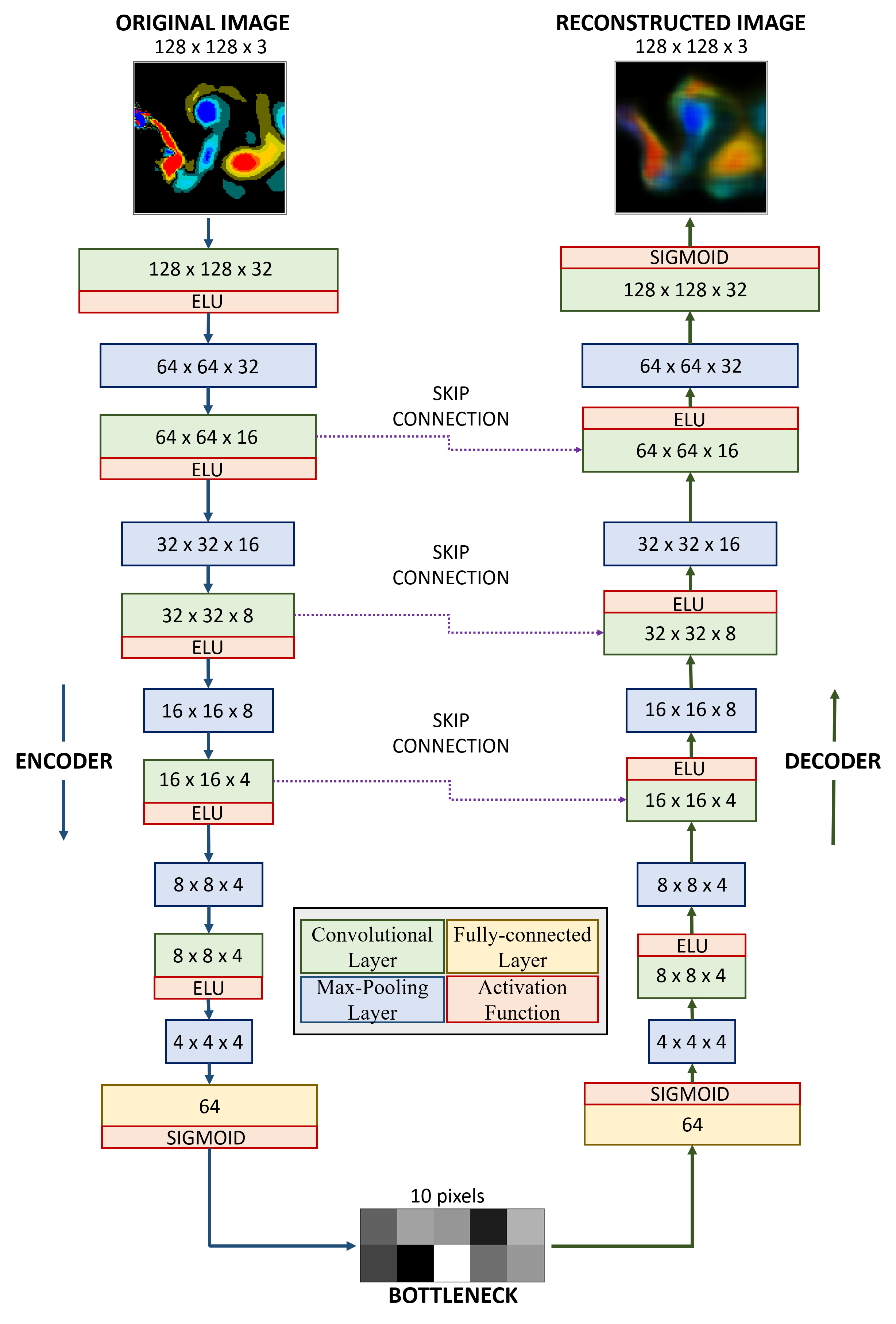}
    \caption*{Tuned convolutional autoencoder architecture.}
    \label{fig:autoSketch}
\end{figure}

\end{document}